\documentclass[10pt, a4paper]{article}
\usepackage{fullpage}
\usepackage{amsmath}
\usepackage{amsfonts}
\usepackage{algorithm}
\usepackage{algorithmicx}

\usepackage[font=small,labelfont=bf]{caption}

\usepackage[
backend=biber,
style=alphabetic,
sorting=ynt
]{biblatex}
\usepackage[colorlinks=true, citecolor=blue]{hyperref}
\usepackage{breakurl}

\title{\textbf{ZK Secret Santa}}

\author{Artem Chystiakov, Kyrylo Riabov}

\date{January 2025}

\begin{document}
\maketitle

\begin{abstract}

This paper proposes a three-step Secret Santa algorithm with setup that leverages Zero Knowledge Proofs (ZKP) to establish gift sender/receiver relations while maintaining the sender's confidentiality. The algorithm maintains a permutational derangement and does not require a central authority to perform successfully. The described approach can be implemented in Solidity provided the integration with a transaction relayer.

\end{abstract}

\section{Introduction}

Everyone loves playing Secret Santa when Christmas comes. But playing the game on-chain has a set of challenges to overcome. 

First, the openness of the Ethereum blockchain does not allow us to perform computations privately. In order to conceal the identities (addresses) of Secret Santa participants, a transaction relayer together with ZKP is used.

Second, since no source of (true) randomness is available on-chain, the choice of gift sender/receiver pair is outsourced to the Secret Santa participants and verified via ZKP, ensuring that no one chooses themselves.

Third, the problem of \texttt{"}double voting\texttt{"} is solved by the concept of nullifers (blinders). Without losing players' confidentiality, the protocol can verify their participation.

\section{Protocol description}

The ZK Secret Santa (ZKSS) protocol is a three-step non-peer-interactive process that requires the involvement of all the game's participants.

At its core, the algorithm relies on several cryptographic primitives to ensure execution correctness and maintain user confidentiality. Let $\mathbb{F}_p$ be a finite field over a prime $p$, and let

\[
\mathsf{hash}(m) \;\rightarrow\; h \;\in\; \mathbb{F}_p
\]

denote a cryptographic hash function that takes an arbitrary message $m$ as input and returns a field element $h$. 

The proof relation is defined as

\[
\mathcal{R} = \{(w, x) \in \mathcal{W} \times \mathcal{X} \;:\; \phi_1(w,x), \phi_2(w,x), \dots, \phi_m(w,x)\},
\]

where $w$ is the witness data, $x$ is the public data, and $\{\phi_1(w,x), \phi_2(w,x), \dots, \phi_m(w,x)\}$ is the set of relations that must be proven simultaneously.

The function $\mathsf{ecrecover}$ \cite{ethereum_yellow_paper} is used to recover the user's $\mathsf{address}$ based on the ECDSA signature $\mathsf{sig}$ and the signed hash $h$.

Finally, consider the Merkle proof $p_i \in \mathbb{F}_p^{(n)}$, which is the list of node values that leads to the root value. The Merkle proof for the element $x$ can be verified by

\[
\mathsf{merkleVerify}(x, p_i, root) \;\rightarrow\; \mathsf{bool}.
\]

\subsection{Setup}

The preliminary \textit{setup} requires all ZKSS participants to register their addresses publicly in a smart contract. The setup has to be done only once, allowing the participants set to be reused in multiple games.

Since setup is an open process, a \textit{lead participant} may be chosen to register all players in a single transaction safely on their behalf.

\begin{algorithm}[H]
\caption{Setup}
\label{alg:setup}
\textbf{Inputs:} 
    \begin{itemize}
        \item \textit{Public:} 
        \begin{itemize}
            \item[] $addresses$ -- ZKSS participants Ethereum addresses
        \end{itemize}
    \end{itemize}
\textbf{Setup process:}
\begin{algorithmic}
    \State \qquad 1. The lead participant calls the register function on the smart contract providing addresses of all the participants.
    \State \qquad 2. The contract stores the addresses in a participants Sparse Merkle Tree (SMT) under a corresponding $\mathsf{index} = \mathsf{hash}(\mathsf{address})$.
\end{algorithmic}
\end{algorithm}

The participants SMT \cite{sparse_merkle_tree} is used throughout the ZKSS to verify that the participant belongs to the initial set of participants.

\subsection{Signature commitment}

The first step, called \textit{signature commitment}, is required to constrain ZKSS participants to use deterministically derived ECDSA signatures. See the \hyperref[ecdsa-non-determinism]{ECDSA non-determinism} security section for the rationale behind this step.

\begin{algorithm}[H]
\caption{Signature commitment}
\label{alg:sigcomm}
\textbf{Inputs:} 
    \begin{itemize}
        \item \textit{Public:} 
        \begin{itemize}
            \item[] $H$ -- a hash of user's ECDSA signature of ($\mathsf{address}\ ||\  \mathsf{eventId}$), where $\mathsf{address}$ is the user's Ethereum address, $\mathsf{eventId}$ is ($\mathsf{contract\ address}\  ||\  \mathsf{nonce}$), and $||$ is a concatenation operation
        \end{itemize}
    \end{itemize}
\textbf{Commitment process:}
\begin{algorithmic}
    \State \qquad 1. The participant signs the message $M$ = ($\mathsf{address}\ ||\  \mathsf{eventId}$) and calculates the signature hash.
    \State \qquad 2. The participant calls the commit function on the smart contract with a hash as a parameter.
    \State \qquad 3. The contract stores the provided hash in a signature commitments SMT under a corresponding $\mathsf{index}$ = $H$.
\end{algorithmic}
\end{algorithm}

Furthermore, during the signature commitment step, a smart contract must verify that $\mathsf{msg.sender}$ belongs to the initial set of ZKSS participants.

\subsection{Gift sender determination}

The second step, called \textit{sender determination}, obligates every ZKSS participant to anonymously add their randomness $r$ to an array of gift senders.

\begin{algorithm}[H]
\caption{Gift sender determination}
\label{alg:determine}
\textbf{Inputs:} 
    \begin{itemize}
        \item \textit{Private:} 
        \begin{itemize}
            \item[] $\mathsf{sig}$ -- user's ECDSA signature of ($\mathsf{address}\ ||\  \mathsf{eventId}$)
            \item[] $\mathsf{address}$ -- user's address
            \item[] $p_p$ -- the Merkle proof for the user's address inclusion
            \item[] $p_c$ -- the Merkle proof for the user's signature commitment inclusion
        \end{itemize}
        \item \textit{Public:} 
        \begin{itemize}
            \item[] $r$ -- user's unique randomness
            \item[] $\mathsf{eventId}$ -- a unique id of a ZKSS game
            \item[] $\mathsf{root_p}$ -- participants SMT root
            \item[] $\mathsf{root_c}$ -- signature commitments SMT root
            \item[] $\mathsf{null_s}$ -- user's nullifier to prevent double registration of randomness
        \end{itemize}
    \end{itemize}
    \textbf{Proving:}
    \begin{algorithmic}
    \State \qquad Generate proof $\pi_e$ for relation: 
        \begin{gather*}
            \mathcal{R}_{e} = \{
                \mathsf{sig}, \mathsf{address}, p_p, p_c, r, \mathsf{eventId}, \mathsf{root_p}, \mathsf{root_c}, \mathsf{null_s}: \\
                \mathsf{null_s} \leftarrow \mathsf{hash}(\mathsf{sig.s}), \\ 
                \mathsf{address} \leftarrow \mathsf{ecrecover}(\mathsf{sig}, (\mathsf{address}\ ||\ \mathsf{eventId})), \\
                \mathsf{merkleVerify}(\mathsf{address}, p_p, \mathsf{root_p}) \rightarrow \mathsf{true}, \\
                \mathsf{merkleVerify}(\mathsf{hash}(\mathsf{sig}), p_c, \mathsf{root_c}) \rightarrow \mathsf{true}, \\ 
                r = r * r
                \}
        \end{gather*}
    \State \qquad The proof $\pi_e$ is verified by the contract. If it's correct and $\mathsf{null_s}$ isn't included in the list of spent nullifiers, the user includes their randomness into the array via relayer. The randomness and nullifier have to be pair-wise accessible.
    \State \qquad The last operation $r = r * r$ in the relation $\mathcal{R}_{e}$ generates additional constraints to \texttt{"}anchor\texttt{"} the $r$ value to ensure the soundness of the protocol.
    \end{algorithmic}
\end{algorithm}

The $r$ must be generated uniquely by every participant and publicly disclosed. However, the relation between $r$ and $\mathsf{participant}$ is hidden in a ZKP with a relayer sending the transaction.

Players are advised to use a 2048-bit RSA public key for the randomness $r$. That is, ZKSS participants uniquely generate RSA private keys (that they must remember) and publish corresponding public keys given $\mathsf{exp}$ = 65537.

The published RSA public keys are then used in the algorithm's third step to encrypt the gift receivers' delivery addresses so that only the respective gift senders can read them.

\subsection{Gift receiver disclosure}

The third step, called \textit{receiver disclosure}, is the final step in the ZKSS algorithm. Afterwards, the Secret Santa distribution will be complete and gift senders may start sending gifts to receivers.

The receiver disclosure step may be carried out without a relayer as a receiver identity ($\mathsf{msg.sender}$) gets revealed regardless.

\begin{algorithm}[H]
\caption{Disclosing gift receiver}
\label{alg:disclose}
\textbf{Inputs:} 
    \begin{itemize}
        \item \textit{Private:} 
        \begin{itemize}
            \item[] $\mathsf{sig}$ -- user's ECDSA signature of ($\mathsf{address}\ ||\ \mathsf{eventId}$)
        \end{itemize}
        \item \textit{Public:} 
        \begin{itemize}
            \item[] $\mathsf{address}$ -- user's address
            \item[] $\mathsf{eventId}$ -- a unique id of a ZKSS game (contract address $||$ nonce)
            \item[] $\mathsf{null_s}$ -- sender's nullifier
        \end{itemize}
    \end{itemize}
    \textbf{Proving:}
    \begin{algorithmic}
    \State \qquad Generate proof $\pi_c$ for relation: 
        \begin{gather*}
            \mathcal{R}_{c} = \{
                \mathsf{sig}, \mathsf{address}, \mathsf{eventId}, \mathsf{null_s}: \\
                \mathsf{null_r} \leftarrow \mathsf{hash}(\mathsf{sig.s}), \\ 
                \mathsf{address} \leftarrow \mathsf{ecrecover}(\mathsf{sig}, (\mathsf{address}\ ||\ \mathsf{eventId})), \\
                \mathsf{null_r} \neq \mathsf{null_s}
                \}
        \end{gather*}
    \State \qquad The proof $\pi_c$ is verified by the contract. If the verification is successful and $\mathsf{null_r}$ does not equal chosen $\mathsf{null_s}$, the receiver's $\mathsf{address}$ is assigned to the corresponding sender's randomness (RSA public key). 
    \State \qquad The nullifiers' inequality must be verified privately to not disclose the receiver's position from the previous step.
    \State \qquad To enforce receiver disclosure uniqueness, a smart contract (publicly) has to maintain the list of unique $\mathsf{msg.senders}$ and verify their belonging to the initial set of ZKSS participants.
    \end{algorithmic}
\end{algorithm}

In case of a collision when multiple receivers simultaneously choose the same sender, one of the transactions must revert and the discarded receiver has to try to disclose themselves again.

Alongside the ZKP, the gift receiver may provide their encrypted (real-world) address where they wish the gift to be delivered. The encryption is performed using the previously provided sender's RSA public key.

The implementation of the ZKSS protocol may add RSA encryption correctness checks to the ZKP verification scheme.

\section{Security}

\subsection{ECDSA non-determinism}
\label{ecdsa-non-determinism}

Without the signature commitment step, it is possible to attack the ZKSS protocol and DoS the game. A dishonest participant may generate non-deterministic ECDSA signatures and bypass the nullifiers protection, occupying all the senders' slots.

That being said, an alternative version of the ZKSS procotol may be proposed using EdDSA signatures. Their deterministic nature allows the signature commitment step to be skipped, the signature commitments SMT removed, and nullifiers constructed directly as $\mathsf{hash}(\mathsf{sig})$, not $\mathsf{hash}(\mathsf{sig.s})$.
    
\subsection{Receiver frontrunning}

There is a possibility of a minor frontrunning attack during the receiver disclosure step of the protocol.

A dishonest gift sender may monitor a transactions mempool to frontrun a receiver (by choosing the same sender) to increase the chances of that receiver choosing the dishonest sender during their next disclosure attempt.

However, this attack works only once (a receiver can only choose a sender once) and is impossible when a dishonest sender is already chosen.

\section{Correctness}

The essential element of ZKSS is to separate its second and third steps. Because a transaction relayer is used during the second step, participants in the third step cannot determine which randomness belongs to which participant. Moreover, participants can cast their randomness only once, which is enforced by nullifers logic. Furthermore, by employing ZKP, we ensure that no participant can manipulate the protocol to send a gift to themselves or choose a particular sender.

To illustrate these concepts, consider the following Secret Santa analogy. Imagine $n$ participants who gather together (\hyperref[alg:setup]{Step 1}).

Each participant securely places a piece of paper containing their randomness into a hat. Everyone adds their respective notes secretly, and no one can observe which note belongs to whom, except for the \texttt{"}transport\texttt{"} mechanism, which corresponds to the relayer in our protocol (\hyperref[alg:determine]{Step 2}). Various technologies, VPNs, etc. can be used to ensure anonymity concerning the relayer, but they go beyond this paper.

Finally, each participant draws exactly one note from the hat, and --- by the \texttt{"}magic\texttt{"} (guaranteed by the ZKP) --- they cannot retrieve their own note (\hyperref[alg:disclose]{Step 3}). After the notes with random numbers are revealed, the corresponding participant associated with a certain number sends a gift to the person who pulled the note out.

Moreover, if the chosen randomness is an RSA public key, the receiver can securely transmit a delivery address to their Santa via RSA encryption.

\vspace{0.3cm}

To summarize, the key assumptions regarding the correctness of the protocol are as follows: 

\begin{enumerate}
    \item Each gift sender will not disclose themselves and the randomness used in the second step of the protocol.
    \item The ECDSA signature must be constructed following RFC 6979 \cite{rfc6979}. Signatures only from the lower half of the curve must be accepted.
    \item The $\mathsf{eventId}$ is unique for every ZKSS game.
    \item The soundness of the protocol derives from the soundness of the underlying ZK proving system. 
    \item If participants provide encrypted payloads, they are responsible for ensuring the correctness of the encrypted data.
\end{enumerate}

\printbibliography

\end{document}